\documentclass[a4paper,12pt]{article}
\pdfoutput=1 

\usepackage{jheppub} 

\usepackage[T1]{fontenc} 
\usepackage{bbm}
\usepackage{float}
\usepackage{mathrsfs}
\usepackage{doi}
\usepackage{natbib, hyperref}
\title{\boldmath On disc-with-hole and disc-with-handle  partition functions in bosonic string theory}

\author[a,b]{Brett Oertel}
\author[a,c]{Omar Shahpo}

\affiliation[a]{Blackett Laboratory Imperial College London London, UK, SW7 2AZ}

\affiliation[b]{Department of Applied Mathematics and Theoretical Physics
University of Cambridge Wilberforce Road, Cambridge, UK, CB3 OWA}

\affiliation[c]{Department of Mathematics 
King's College London
The Strand, London, UK, WC2R 2LS}

\emailAdd{bo291@cam.ac.uk}
\emailAdd{omar.shahpo@kcl.ac.uk}

\abstract{Higher genus  partition functions of   string world sheets  with boundaries 
are 
relevant, e.g.   for  computation of 
 quantum corrections to Wilson loop expectation values.  As a preparation for a possible 
 study  of   strings in curved space  like AdS  
  here we  consider  examples of 
 genus one partition functions of string world-sheets ending on a circle in the bosonic  string theory in flat space. 
 We begin with  the partition function for  annular topology, writing it as an integral over the modulus of the annulus. In the process, we compute the determinant of the Laplacian on the annulus for  Dirichlet-Dirichlet and Dirichlet-Neumann boundary conditions. We then consider  the case of the disc-with-handle topology using the gluing method.  We first write the partition function using a Schottky parameterisation of the moduli space and then as an integral over the period matrix.
 
\vfill
\vspace{8cm}

\begin{center}
    Presented here is joint work done by the authors towards theses submitted in partial fulfillment of the requirements for the MSci degree in Physics with Theoretical Physics of Imperial College London.
\end{center} 
            
\vfill

 }

\begin{document} 
\maketitle
\flushbottom

\section{Introduction}

The correspondence between Wilson loops (WLs) in  large $N$  $\mathcal{N}=4$ SYM gauge 
 theory and partition functions of $AdS_5\times S^5$ string world-sheets ending on a fixed curve has been heavily studied since \cite{PhysRevLett.80.4859, PhysRevD.60.125006}.
A systematic  discussion of leading  string inverse tension 
  quantum  corrections on disc  topology appeared  in 
\cite{Drukker:2000ep} (following \cite{Kallosh:1998ji,Forste:1999qn}).
Some recent papers   discussing matching between   string and   strong-coupling 
gauge theory results for circular WL include \cite{Medina-Rincon:2018wjs,Giombi:2020mhz}.

 The next step is matching  non-planar $1/N$ corrections to Wilson loops  with 
 higher genus  string partition functions. The former are known to all orders 
 for the circular WL case   \cite{Drukker:2000rr} and a discussion of possible matching 
 was initiated in \cite{Giombi:2020mhz}.
 Before trying to address the    $AdS_5\times S^5$  string   case it is useful  to review what 
can be said  explicitly   about the  corresponding  corrections  in the  bosonic string theory in flat 
   target space.  The  $1/N$  correction to the  circular WL expectation value\footnote{It is absent in 
   adjoint  $SU(N)$  gauge theory   but  
   may be interpreted as due to a fermion loop in fundamental representation  in more general cases involving    orientifolds, see, e.g.,    \cite{Giombi:2020kvo}.}
corresponds to the string  partition function on the annulus, i.e. to adding a hole (with Neumann  b.c.)  to the disc
(with Dirichlet b.c.)  and integrating over the size of the hole. 
Similarly, the $1/N^2$   correction to the WL corresponds to the string partition function on a surface with topology of a   disc  with a  handle. 


We begin in Section 2 by considering the case of the partition function of the world-sheet of annular topology with Dirichlet condition on one circle and Neumann on  the other.
We  first compute the determinant of the Laplacian on the annulus for the case of Dirichlet-Dirichlet boundary conditions, which is a simple modification of the determinant of the Laplacian on the torus,
and then   for the case of the Dirichlet-Neumann conditions. The result in the latter case 
 agrees with that  
  found in  \cite{Graham:2018ibr}  by a different method. 

In Section 3  we tackle the case of the partition function on the  disc with a handle. This is done 
following the  approach of  \cite{BLAU1988285}, by a gluing a disc with a handle to itself, obtaining a double torus,  and then parameterising the moduli of the disc-with-handle using the moduli of the double torus. We compute the corresponding partition function using two different  parametrisations of the moduli space: 
 first using the Schottky parameters and then using the period matrix.

\section{The Annulus}


We work with the Polyakov action as given in \cite{Alvarez:1982zi}:
\begin{equation}
    I'[X,g] = \frac{1}{2} T \int_{\Sigma}d^{2}\xi\sqrt{g}g^{ab}\delta_{\mu\nu}\partial_{a}X^{\mu}\partial_{b}X^{\nu} + \phi_0   \chi + \mu^2 \int_{\Sigma}d^{2}\xi \sqrt{g} + I_{\mathrm{b}}.
\end{equation}
Here $\Sigma$ is a two-dimensional compact Riemannian manifold (in this section with the topology of an annulus), the $X^{\mu}$ represent 26 real scalar fields on $\Sigma$ which determine its embedding into $\mathbb{R}^{26}$, $g$ is the Riemannian metric on $\Sigma$ and $\chi$ is the Euler number of $\Sigma$. 
$T=\frac{1}{ 2 \pi \alpha'}$ is string tension. 

We shall  assume that the constants $\mu$ and the counter terms $I_b$ are chosen so that the conformal anomaly cancels in 26 dimensions \cite{Alvarez:1982zi}, and we drop these terms. The Euler number term provides the weighting of the string partition function with string coupling that we shall suppress.
Thus in what follows we shall use 
\begin{equation}
    I[X,g] = \frac{1}{2}T \int_{\Sigma}d^{2}\xi\sqrt{g}g^{ab}\delta_{\mu\nu}\partial_{a}X^{\mu}\partial_{b}X^{\nu}.
\end{equation}
The string world-sheet partition function is \cite{Alvarez:1982zi}
\begin{equation}
    W = \frac{1}{V_{\mathrm{GC}\times\mathrm{Weyl}}}\int DgDX\  e^{-I[X,g]} 
\end{equation}
The integral is over all embeddings $X^{\mu}$ of $\Sigma$ and all intrinsic metrics $g$ on $\Sigma$ (both must satisfy boundary conditions to be discussed). We divide by the volume of the classical gauge group $V_{\mathrm{GC}\times\mathrm{Weyl}}$ where $\mathrm{GC}\times\mathrm{Weyl}$ is the direct product of the general coordinate transformation and Weyl groups.

\subsection{The parametrisation and boundary conditions}
To parametrise the annulus we use a strip in the complex plane:
\begin{equation}
    \{\xi^i\}, \hspace{0.5cm} \xi^1,\xi^2 \in \mathbb{R}, \hspace{0.5cm} 0 \leq \xi^2 \leq1.
\end{equation}
This parametrises the annulus in the complex plane given by
\begin{equation}
    q\leq|z|\leq1, \hspace{0.5cm} 0<q<1 \hspace{0.5cm} z \in \mathbb{C} 
\end{equation}
using the map
\begin{equation}
    z = \exp(2\pi i (\xi^1+i\tau \xi^2)), \hspace{0.5cm} \tau \equiv -\frac{1}{2\pi}\ln q .
\end{equation}
Under this map, which is periodic with period 1 in the $\xi^1$ direction, the top boundary ($\xi^2 = 1$) of the strip is mapped to the inner boundary of the annulus, while the bottom boundary ($\xi^2 = 0$) is mapped to the outer boundary of the annulus. Since we are interested in the path integral over the annulus as a one-loop correction to a phenomenological model of the Wilson loop, we will use `modified Dirichlet' \cite{Alvarez:1982zi} conditions for the embedding functions $X^{\mu}$ on the outer boundary. After gauge fixing, these become Dirichlet boundary conditions. On the other hand, the inner boundary is not mapped to a fixed curve, and we simply choose Neumann boundary conditions on it. Next, the $X^{\mu}$ and $g$ are obviously required to be periodic in the $\xi^1$ direction:
\begin{equation}
\begin{aligned}
&X^{\mu}(\xi^1 + 1, \xi^2) = X^{\mu}(\xi^1,\xi^2),\\
&g_{ab}(\xi^1 + 1, \xi^2) = g_{ab}(\xi^1,\xi^2).
\end{aligned}
\end{equation}
On the upper boundary and lower boundaries $g_{ab}$ must satisfy  \cite{Alvarez:1982zi}
\begin{equation}
    g_{ab} = \partial_{a}X^{\mu}\partial_{b}X_{\mu},
\end{equation}
 The boundary conditions needed to compute the determinant of the Laplacian are discussed below.
 
\subsection{The path integral}
To perform the $X^{\mu}$ integration one redefines the integration variable as an expansion about the classical solution:
\begin{equation}
    X^{\mu} = X^{\mu}_{cl} + y^{\mu}.
\end{equation}
Here $X^{\mu}_{cl}$ is the solution to the classical equation of motion obtained by varying the action. Note that this solution depends on $g$, thus the answer will be a function of the metric and remains inside the integral when we integrate over the metric. Then, the $X^\mu $ integration  gives\footnote{We  set the string tension $T$ 
 to 1   below as it can be scaled out of the partition function.}
\begin{equation}
\begin{aligned}
    W_{g} &\equiv  \int DX\ e^{- I[X,g]} 
     = e^{-I[X_{cl},g]}\left(\mathrm{det} (\Delta_{g})\right)^{-13},
\end{aligned}
\end{equation}
where $\Delta_{g}$ is the ordinary Laplacian differential operator, i.e.
\begin{equation}
    \Delta_g \equiv -\frac{1}{\sqrt{g}}\partial_{a}\sqrt{g}g^{ab}\partial_{b}. 
\end{equation}
The determinant of $\Delta_{g}$ is taken with Dirichlet boundary conditions. Note that $X_{cl}$ depends on $g$ only through the moduli variables since $I$ is invariant under diffeomorphisms and conformal transformations. Our final result for the partition function will be in the form of an integral over these moduli. To perform the integration over $g$ we first note that any metric on the annulus can be taken via a combined general coordinate and Weyl transformation (respecting the above boundary conditions) to a metric of the form (see, e.g., \cite{Rodrigues:1986us})
\begin{equation}
    g_{ab} = (d\xi^1)^2+(\tau d\xi^2)^2.
\end{equation}
Here $\tau = -\frac{1}{2\pi}\ln q$ is the Teichmuller parameter.
The integral over metrics then  reduces (in $D=26$) to the integral over $\tau$
\begin{equation}
\begin{aligned}
    W = &\frac{1}{2}\int_{0}^{\infty} d\tau \frac{1}{\sqrt{\tau} } \left(\frac{\det^2 <T,\psi>}{\det <\psi,\psi>}\right)^{\frac{1}{2}}\left(\mathrm{det}' P^{\dagger}_{1}P_{1}\right)^{\frac{1}{2}}\\
    & \times \left(\mathrm{det} (\Delta_{g})\right)^{-13}e^{-I[X_{cl},g]}.
\end{aligned}
\end{equation}
where we have used the following definitions:
\begin{equation}
    \begin{aligned}
    (T_{ab}) &\equiv (\frac{\partial g_{ab}}{\partial \tau} - \frac{1}{2}g_{ab}g^{cd}\frac{\partial g_{cd}}{\partial \tau})\\
     (P_{1}\delta v)_{ab} &\equiv \nabla_{a}\delta v_{b} + \nabla_{b}\delta v_{a} - \nabla_{c}\delta v^{c}g_{ab}\\
    (P_1^{\dagger}\alpha)_n &\equiv - 2\nabla^{m}\alpha_{mn}\ . 
    \end{aligned}
\end{equation}
 $P_1$ acts on two-dimensional vectors (diffeomorphism generators) $\delta v$ and $P_1^{\dagger}$ acts on rank-2 traceless tensors $\alpha$. The norm on rank-2 traceless tensors is defined by: 
\begin{equation}
\label{norm}
    <A,B>_{g} = \int_{\Sigma} d\sigma^1 d\sigma^2 \sqrt{g(\sigma)}\ g^{mi}(\sigma) g^{nj}(\sigma)\ A_{mn}(\sigma)B_{ij}(\sigma)
\end{equation}
and $\psi$ is any element in $\mathrm{Ker}(P_{1}^{\dagger})$. The Riemann-Roch theorem  implies \cite{Alvarez:1982zi}:
\begin{equation}
    \mathrm{dim} \ \mathrm{Ker}(P_1) -  \mathrm{dim} \ \mathrm{Ker}(P_{1}^{\dagger}) = 3\chi(\Sigma) = 0. 
\end{equation}
From this we see that the  conformal Killing vectors (CKV), i.e.   vectors in  in $\mathrm{Ker}(P_1)$, 
 on the annulus form a one-dimensional real vector space  (corresponding  to rigid translation in the $\xi^{1}$ direction). Also note that the mapping class group is isomorphic to $\mathbb{Z}_{2}$ and is given by the coordinate transformation \cite{Rodrigues:1986us}
\begin{equation}
    \xi^1 \rightarrow -\xi^1, \hspace{0.5cm} \xi^2 \rightarrow 1-\xi^2.
\end{equation}
The volume of the mapping class group is therefore $2$, and this is where the factor of $\frac{1}{2}$ in front of the path integral comes from.

\subsection{The Determinant and the Weil-Petersson Measure}
We can calculate the Weil-Petersson measure directly. We have
 \begin{equation}
     (T_{ab}) \equiv (\frac{\partial g_{ab}}{\partial \tau} - \frac{1}{2}g_{ab}g^{cd}\frac{\partial g_{cd}}{\partial \tau}) =
    \frac{1}{\tau}    
    \begin{pmatrix}
    -1 & 0 \\
    0 & \tau^2
    \end{pmatrix}.
\end{equation}
The boundary conditions and the symmetric and traceless requirement on matrices in $\mathrm{Ker}(P_1^{\dagger}$) mean we must choose
\begin{equation}
    \psi = 
    \begin{pmatrix}
    1 & 0 \\
    0 & -\tau^2
    \end{pmatrix}.
\end{equation}
A direct calculation then gives 
\begin{equation}
    \left(\frac{\det^2 <T,\psi>}{\det <\psi,\psi>}\right)^{\frac{1}{2}} = \sqrt{\frac{2}{\tau}}.
\end{equation}
All that is left is to calculate the determinants of the Laplacians. Firstly, 
\begin{equation}
\begin{aligned}
    (P_1^{\dagger}P_1\delta v)_a &= -2
    \begin{pmatrix}
    \nabla_g^2 & 0 \\
    0 & \nabla_g^2
    \end{pmatrix}
    \begin{pmatrix}
    \delta v_1\\
    \delta v_2
    \end{pmatrix}.
\end{aligned}
\end{equation}
We have chosen boundary conditions such that changes in the metric do not change the normal direction at the boundary. This imposes Neumann boundary conditions on $\delta v_1$ and Dirichlet on $\delta v_2$ \cite{Rodrigues:1986us}. Hence,
\begin{equation}
    \mathrm{det}'(P^{\dagger}_1P_1) = \mathrm{det}'_N(-2\nabla_g^2)\ \mathrm{det}'_D(-2\nabla_g^2).
\end{equation}
Furthermore, the factor $ (\mathrm{det}'(\Delta_g))^{-13}$ which arises due to the $X$-integration must be taken with Dirichlet boundary conditions at the lower edge ($\xi^2 = 0$) and Neumann at the upper edge ($\xi^2=1$). Now we calculate each determinant individually. First, the operator in the determinant
\begin{equation}
    \mathrm{det}'_N(-2\nabla_g^2) = \mathrm{det}'_N(-2(\partial_1^2+\frac{1}{\tau^2}\partial_2^2))
\end{equation}
has eigenvectors
\begin{equation}
    \cos (2\pi m \xi^1) \cos(\pi n \xi^2 ), \hspace{0.5cm} m,n \in \mathbb{Z} \hspace{0.5cm} m,n \geq 0, \hspace{0.5cm} (m,n) \neq (0,0)
\end{equation}
and the corresponding eigenvalues
\begin{equation}
\begin{aligned}
    \lambda_{mn} &= \frac{8\pi^2}{\tau'^2}|\tau''m+n|^2, \hspace{0.5cm} \tau'\equiv 2\tau, \tau'' \equiv i\tau'.
\end{aligned}
\end{equation}
These are identical to the eigenvalues encountered in the one-loop closed string (torus) calculation in the case that the torus Teichmuller parameter is purely imaginary. Such a determinant is calculated, for example, in \cite{Nakahara:2003nw}  using the Eisenstein series. This method can be immediately adapted to obtain: 
\begin{equation}
    \mathrm{det}'_N(-2\nabla_g^2) = \frac{1}{\sqrt{2}}2\tau|\eta (2i\tau)|^2 = \sqrt{2}\tau|\eta (2i\tau)|^2.
\end{equation}
Next, the operator in the  determinant
\begin{equation}
    \mathrm{det}'_D(-2\nabla_g^2) = \mathrm{det}'_D(-2(\partial_1^2+\frac{1}{\tau^2}\partial_2^2))
\end{equation}
has eigenvectors
\begin{equation}
    \cos (2\pi m \xi^1) \sin(\pi n \xi^2 ), \hspace{0.5cm} m,n \in \mathbb{Z}, \hspace{0.5cm} n > 0, \hspace{0.5cm} m \geq 0,
\end{equation}
and identical eigenvalues
\begin{equation}
\begin{aligned}
    \lambda_{mn} &= \frac{8\pi^2}{\tau'^2}|\tau''m+n|^2, \hspace{0.5cm} \tau'' \equiv i2\tau.
\end{aligned}
\end{equation}
The only difference is that we allow $n=0$ so long as $m\neq 0$ in the Neumann case. Hence we obtain
\begin{equation}
    \mathrm{det}'_N(-2\nabla_g^2) = \mathrm{det}'_D(-2\nabla_g^2) \prod_{m \neq 0}(2\pi^2m^2).
\end{equation}
We may regularize this factor as follows:
\begin{equation}
    \prod_{m \geq 0}(2\pi^2m^2) = \exp (-\zeta'_{\Delta}(0)),
\end{equation}
where
\begin{equation}
     \zeta_{\Delta}(s) \equiv \sum_{m\geq 1}(2\pi m)^{-2s} = \frac{1}{(2\pi)^{-2s}}\zeta_{R}(2s).
\end{equation}
Then 
\begin{equation}
    -\zeta'_{\Delta}(0) = (2\zeta'_{R}(0) - 2\log (2\pi)\zeta_{R}(0)) = 0,
\end{equation}
and hence the factor regularizes to $1$. 

Lastly, we need to find the determinant of the Laplacian on the annulus $\mathrm{det}(\Delta_g)$ where the inner boundary has Neumann boundary conditions, and the outer boundary has Dirichlet boundary conditions. We have
\begin{equation}
    \mathrm{det} (\Delta_g) = \mathrm{det}(-(\partial_1^2+\frac{1}{\tau^2}\partial_2^2)).
\end{equation}
This has eigenvectors
\begin{equation}
    \cos (2\pi m \xi^1) \sin(\pi (n+\frac{1}{2}) \xi^2 ), \hspace{0.5cm} m,n \in \mathbb{Z}, \hspace{0.5cm} m,n \geq 0
\end{equation}
with eigenvalues
\begin{equation}
\begin{aligned}
    \lambda_{mn} &= \Big((2\pi m)^2+ \frac{1}{\tau ^2}(\pi (n+\frac{1}{2})\Big)^2.\\
\end{aligned}
\end{equation}
To calculate this determinant we will  need to use a different method, relying on contour integrals. Our method is similar to a spectral method used in \cite{Kirsten:1999qjn}. Consider the initial value problem
\begin{equation}
    \begin{aligned}
    \phi_{m,\mu} ''(\xi ^2) = -\tau^2(\mu^2 - 4\pi^2m^2)\phi_{m,\mu}(\xi^2), \hspace{0.5cm} \phi_{m,\mu}(0) = 0, \hspace{0.5cm} \phi_{m,\mu}'(0) = 1.
    \end{aligned}
\end{equation}
This has the unique solution
\begin{equation}
    \phi_{m,\mu}(\xi^2) = \frac{1}{\tau\sqrt{\mu^2-4\pi^2m^2}}\sin((\tau\sqrt{\mu^2-4\pi^2m^2})\xi^2).
\end{equation}
In particular, for a given $m$ we will view $\phi'_{m, \mu}(1)$ as a meromorphic function of $\mu$:
\begin{equation}
    \phi'_{m,\mu}(1) = \cos(\tau\sqrt{\mu^2-4\pi^2m^2}).
\end{equation}
This function has zeros exactly when 
\begin{equation}
    \mu^2= (4\pi^2m^2 + \frac{1}{\tau^2}(n+\frac{1}{2})^2) \equiv \lambda_{mn},
\end{equation}
i.e. exactly when $\mu^2$ is an eigenvalue of $\Delta_{g}$. Hence, we will consider the function 
\begin{equation}
    \frac{d}{d\mu}\log(\phi'_{m,\mu}(1)) = \frac{ \frac{d}{d\mu}\phi'_{m,\mu}(1)}{\phi'_{m,\mu}(1)}
\end{equation}
which clearly has a simple pole at each $\mu^2 = \lambda_{mn}$ if we consider the expansion of $\phi'_{m,\mu}(1)$. Hence, we may write the zeta function as 
\begin{equation}
    \zeta_{\Delta}(s) = \frac{1}{2\pi i}\int_{\Lambda}d\mu \mu^{-2s}\frac{d}{d\mu}\log(\phi'_{0,\mu}(1)) + \sum_{m = 1}\frac{1}{2\pi i}\int_{\Lambda}d\mu \mu^{-2s}\frac{d}{d\mu}\log(\phi'_{m,\mu}(1))
\end{equation}
where $\Lambda$ is a contour enclosing all the positive simple poles, which all lie on the real axis. Now we choose the contour $\Lambda$ to be parametrised as 
\begin{equation}
    \Lambda(k) = \epsilon + ik, \hspace{0.5cm} \epsilon, k \in \mathbb{R}
\end{equation}
where $\epsilon$ is a small positive number and $k$ runs from $-\infty$ to $\infty$. Substituting this into the integral and taking the limit of $\epsilon$ going to zero we obtain
\begin{equation}
    \begin{aligned}
    \zeta_{\Delta}(s) = &\frac{\sin(\pi s)}{\pi}\int_{0
    }^{\infty}dk k^{-2s}\frac{d}{dk}\log(\phi'_{0,ik}(1))\\
    & + \sum_{m = 1}m^{-2s}\frac{\sin(\pi s)}{\pi}\int_{0
    }^{\infty}dk k^{-2s}\frac{d}{dk}\log(\phi'_{m,imk}(1)),
    \end{aligned}
\end{equation}
where
\begin{equation}
\begin{aligned}
    &\phi'_{0,ik}(1) = \cos(\tau ik) = \cosh(\tau k),\\
    &\phi'_{m,imk}(1) = \cos(\tau\sqrt{-m^2k^2-4\pi^2m^2}) = \cosh(\tau\sqrt{m^2k^2+4\pi^2m^2}).
\end{aligned}
\end{equation}
Now we will regularise these integrals separately. 
\subsubsection{The m=0 term}
First, note that 
\begin{equation}
    \frac{d}{dk}\log(\phi'_{0,ik}(1)) = \frac{d}{dk}\log(\cosh(\tau k)) = \tau \tanh(\tau k).
\end{equation}
This has the following asymptotic expansion:
\begin{equation}
    \tau \tanh(\tau k) = \tau + \mathcal{O}(\exp(-2\tau k)).
\end{equation}
Hence, we will write
\begin{equation}
    \frac{\sin(\pi s)}{\pi}\int_{0
    }^{\infty}dk k^{-2s}(\tau \tanh(\tau k)-\tau + \tau) = Z_{0}(s) + A_{0}(s),
\end{equation}
where
\begin{equation}
    \begin{aligned}
    &Z_{0}(s) \equiv \frac{\sin(\pi s)}{\pi}\left(\int_{0
    }^{1}dk k^{-2s}(\tau \tanh(\tau k))+\int_{1
    }^{\infty}dk k^{-2s}(\tau \tanh(\tau k)-\tau)\right),\\
    &A_{0}(s) \equiv \frac{\sin(\pi s)}{\pi}\int_{1
    }^{\infty}dk k^{-2s}\tau.
    \end{aligned}
\end{equation}
$Z_{0}(s)$ is not divergent at $s=0$ and we find
\begin{equation}
\begin{aligned}
    Z'_{0}(0) &= \int_{0}^{1}dk\frac{d}{dk}(\log ( \cosh (\tau k)))+\int_{1}^{\infty}dk\frac{d}{dk}(\log ( \cosh (\tau k)) - \tau k)\\
    & = -\log (2)+\tau.
\end{aligned}
\end{equation}
On the other hand, we may regularise $A_{0}(s)$ as 
\begin{equation}
\begin{aligned}
&A_{0}(s) = \frac{\tau \sin (\pi s)}{\pi(2s-1)}
\implies{} & A_{0}'(0) = -\tau.
\end{aligned}
\end{equation}
In total, for the $m=0$ term we obtain
\begin{equation}
    \left.\frac{\sin(\pi s)}{\pi}\int_{0
    }^{\infty}dk k^{-2s}\frac{d}{dk}\log(\phi'_{0,ik}(1))\right|_{s=0} = -\log(2).
\end{equation}
\subsubsection{The m > 0 terms}
Again, we have 
\begin{equation}
    \begin{aligned}
    &\frac{d}{dk}\log(\phi'_{m,ik}(1))) = \frac{d}{dk}\log(\cosh(\tau m\sqrt{k^2+4\pi^2})),\\
    \end{aligned}
\end{equation}
and
\begin{equation}
    \begin{aligned}
    \sum_{m=1}m^{-2s}\frac{\sin(\pi s)}{\pi}\int_{0}^{\infty}dk k^{-2s}\frac{d}{dk}\log(\phi'_{m,imk}(1)) = Z_{m}(s)+A_{m}(s).
    \end{aligned}
\end{equation}
We have defined
\begin{equation}
    \begin{aligned}
    Z_{m}(s) \equiv &\sum_{m=1}\left(m^{-2s}\frac{\sin(\pi s)}{\pi}\int_{0}^{1}dkk^{-2s}\frac{d}{dk}(\log ( \cosh(\tau m\sqrt{k^2+4\pi^2}))\right.\\
    &+m^{-2s}\frac{\sin(\pi s)}{\pi}\int_{1}^{\infty}dkk^{-2s}\frac{d}{dk}(\log ( \cosh (\tau m\sqrt{k^2+4\pi^2})-m \tau k)\\
    &-\left.m^{-2s}\frac{\sin(\pi s)}{\pi}\int_{1}^{\infty}(1 -2\pi)\tau m \exp(-k+1)\right),\\
    A_{m}(s) \equiv &\frac{\tau\sin(\pi s)}{\pi}(\frac{1}{2s-1}+(1-2\pi))\zeta_{R}(2s-1).
    \end{aligned}
\end{equation}
The terms that we have added and subtracted are specifically chosen so that $Z_{m}(s)$ is well defined in a neighbourhood of $s=0$. Hence we simply differentiate and set $s=0$, obtaining
\begin{equation}
\begin{aligned}
    Z_{m}'(0) = -\sum_{m=1}\log(1+\exp(-4\pi\tau m)).
\end{aligned}
\end{equation}
Lastly, the divergent piece $A_{m}(s)$ can now be regularized in the obvious manner, by analytically continuing the Riemann zeta function:
\begin{equation}
    A'_{m}(0) = \frac{\pi \tau}{6}.
\end{equation}
Hence, we may combine all results to obtain
\begin{equation}
\begin{aligned}
    &\zeta_{\Delta}'(0) = -\log(2)+\frac{\pi\tau}{6}-\sum_{m=1}\log(1+\exp(-4\pi\tau m)),\\
    &\textrm{det}(\Delta_g) = 2\exp(\frac{-\pi \tau}{6})\prod_{m=1}(1+\exp(-4\pi\tau m)).
\end{aligned}
\end{equation}
The final expression for the partition function then reads
\begin{equation}
    W = \int_{0}^{\infty}d\tau\ |\eta(2i\tau)|^{2}\left(\frac{1}{2}\exp(\frac{\pi \tau}{6})\prod_{m=1}(1+\exp(-4\pi\tau m))^{-1}\right)^{13}e^{-\pi(1-q^2)}.
\end{equation}

Note that since we parametrised the annulus by a rectangular strip, we can check this result by looking for calculations of the determinant of the Laplacian on the rectangle with appropriate boundary conditions. Indeed, this is done using a different method in \cite{Brink:1973rq}, and we find agreement between their result and ours. Note also that here we have used the fact that the classical action $I[X_{cl}, g]$ is simply the minimal area of the annulus as a function of the modulus parameter: 
\begin{equation}
    I[X_{cl},g] = \pi (1-q^2). 
\end{equation}

 The integral over $\tau$ is formal as it is divergent for large $\tau$ corresponding  to small $q= e^{-2\pi \tau}$, i.e. the integral is divergent when the inner radius of the annulus 
 goes to zero. 

\section{The disc-with-handle}
Let us now  compute the partition function for the case where $\Sigma$ is topologically a disc-with-handle, with boundary again given by a fixed circle in spacetime. 
Here again the partition function is  given by 
\begin{equation}
    W = \frac{1}{V_{\mathrm{GC}\times\mathrm{Weyl}}}\int DgDX \exp\left(-\frac{1}{2}\int_{\Sigma}d^{2}\xi\sqrt{g}g^{ab}\delta_{\mu\nu}\partial_{a}X^{\mu}\partial_{b}X^{\nu}\right) \ .
\end{equation}
There are no CKV on the disc-with-handle  so that we get in general 
  \cite{Alvarez:1982zi}:
\begin{equation}
    W = \int_{\mathscr{F}} d\mu_{\mathrm{WP}}\sqrt{\det (P_1^{\dagger}P_1)}\left(\det(\Delta)_{g}\right)^{-13}\exp(-I(X_{cl},g)).
\end{equation}
Here $d\mu_{\mathrm{WP}}$ is the Weil-Petersson measure on the moduli space of $\Sigma$ and the integration is over a fundamental region $\mathscr{F}$.  $P_1^{\dagger}P_1$ and $\Delta$ are the vector and scalar Laplacians on $\Sigma$ respectively, $g$ is a choice of gauge-fixed 
 metric, and $I_{\mathrm{cl}}(X,g)$ is the classical action. The boundary conditions are the same as the outer boundary in the annulus case: $\det(\Delta)_{g}$ is calculated using Dirichlet boundary conditions and $\det (P_1^{\dagger}P_1)$ using mixed boundary conditions.

\subsection{Schottky doubles}
To evaluate $W$  we need to parametrise the moduli space of $\Sigma$ and evaluate determinants on $\Sigma$. In fact, it will be easier to work with the Schottky double of $\Sigma$, which will be a surface without boundary denoted $\Sigma^2$; we will closely follow the approach outlined in \cite{BLAU1988285}. $\Sigma^2$ is topologically constructed by attaching two copies of $\Sigma$, denoted $\Sigma$ and $\Sigma^*$, with opposite orientation along the boundaries. Since $\Sigma$ is a disc-with-handle, $\Sigma^2$ is therefore a double-torus. The key point is that the moduli space and determinants on $\Sigma$ can then be related to those on $\Sigma^2$ in a manner we will now describe. First, recall that the moduli and Teichmuller spaces on $\Sigma$ can be defined as \cite{BLAU1988285}
\begin{equation}
    \mathscr{M}(\Sigma) \equiv \frac{C(\Sigma)}{\mathrm{Diff}(\Sigma)}, \hspace{0.5cm} \mathscr{T}(\Sigma) \equiv \frac{C(\Sigma)}{\mathrm{Diff_0}(\Sigma)}, 
\end{equation}
where $\mathrm{Diff(\Sigma)}$ is the group of orientation preserving diffeomorphisms on $\Sigma$, $\mathrm{Diff}_0 (\Sigma)$ is the subgroup connected to the identity, and $C(\Sigma)$ is the set of almost complex structures on $\Sigma$  (which is in bijection with the set of classes of metrics where metrics in a given class differ only by a Weyl transformation). Now, there exists a natural choice of involution $I$ on the constructed topological space $\Sigma^2$ such that
\begin{equation}
    I^2 = 1, \hspace{0.5cm} I(\Sigma) = \Sigma^*, \hspace{0.5cm} I(\Sigma^*) = \Sigma, \hspace{0.5cm} I|_{\partial\Sigma} = I|_{\partial\Sigma^*} = 1.
\end{equation}
Furthermore, if the surface $\Sigma$ comes with a choice of almost complex structure $J$, then we may define an almost complex structure $\Tilde{J}$ on $\Sigma^2$ as \cite{BLAU1988285} 
\begin{equation}
    \Tilde{J}_{p} = J_{p} \ \textrm{if} \ p \in \Sigma, \hspace{0.5cm} \Tilde{J}_{p} = -(I_{*}\circ J_{p} \circ I_{*})_{p} \ \textrm{if} \ p \in \Sigma^*.
\end{equation}
Then, $I$ is in addition anti-conformal. On the other hand, an arbitrary almost complex structure $J$ on $\Sigma^2$ may or may not make $\Sigma^2$ the double of a surface; it will be considered the double of the surface $\Sigma$ if $J$ is such that the corresponding involution $I$ is anti-conformal. In particular, in \cite{BLAU1988285} the following important facts are noted: any metric on $\Sigma$ is equivalent in $\mathscr{T}(\Sigma)$ to a metric which extends smoothly to a metric on $\Sigma^2$ which is invariant under $I$, and any metric on $\Sigma^2$ invariant under $I$ (up to a diffeomorphism and Weyl transformation) is equivalent in $\mathscr{T}(\Sigma^2)$ to a metric exactly invariant under $I$. Hence, $\mathscr{T}(\Sigma)$ is a natural subset of $\mathscr{T}(\Sigma^2)$. In fact, we will therefore be able to write the partition function over $\Sigma$ in terms of objects defined on $\Sigma^2$. In particular, the determinants of interest on $\Sigma$ can be expressed in terms of determinants on $\Sigma^2$ using the following formulas \cite{BLAU1988285}:
\begin{equation}
    \begin{aligned}
    &\det (\Delta)_{g, \Sigma} = \left(\frac{\det'(\Delta)_{g, \Sigma^2}}{\int_{\Sigma^2}\sqrt{g}}\right)^{\frac{1}{2}}(R_{\Sigma^2, I}(J))^{\frac{1}{2}},\\
    & \det(P_1^{\dagger}P_1)_{g, \Sigma} = (\det(P_1^{\dagger}P_1)_{g, \Sigma^2})^{\frac{1}{2}}.
    \end{aligned}
\end{equation}
Here $ R_{\Sigma^2, I}(J)$ is   \cite{BLAU1988285}
\begin{equation}
     R_{\Sigma^2, I}(J) = \det[(1 + \Gamma) \textrm{Im} \ \tau + (1-\Gamma)(\textrm{Im}\ \tau)^{-1}]
\end{equation}
where $\tau$ and $\Gamma$ are defined as follows. 
First, we define a canonical homology basis of $A$ cycles and $B$ cycles on our Schottky double $\Sigma^{2}$. In general, the involution $I$ will take $A$ cycles to $A$ cycles and $B$ cycles to $B$ cycles. 
Then we define $\Gamma$ as the matrix corresponding to how the involution pushes forward A cycles to other A cycles:
\begin{equation}
    I_{*}A_{i} = \Gamma_{ij}A_{j}.
\end{equation}
In addition, if we let $\{\omega_{i}\}$ be a basis of holomorphic differentials on the Schottky double, normalised to the A-cycles in the sense that 
\begin{equation}
    \int_{A_{i}}\omega_{j} = \delta_{ij},
\end{equation}
then we define the `period matrix' $\tau$ as 
\begin{equation}
    \int_{B_i}\omega_{j} = \tau_{ij}.
\end{equation}
Now we turn to the Weil-Petersson measure $d\mu_{\mathrm{WP}}$ on $\Sigma$. Let $\{\mu_{i}\}$ and $\{S_{i}\}$ be bases for the spaces of Beltrami and quadratic differentials on $\Sigma^2$ which are even under the involution $I$. Each Beltrami differential $\mu_{i}$ corresponds to a tangent vector in the Teichmuller space which we write as $\frac{d}{dm_{i}}$. Then the Weil-Petersson measure on $\Sigma$ can be written as \cite{BLAU1988285} 
\begin{equation}
    d\mu_{\mathrm{WP}, \Sigma} = \frac{\det <S_i, \mu_j>}{(\det<S_i,S_j>)^{\frac{1}{2}}}\prod_{i=1}^{3}dm_{i}.
\end{equation}
This measure is on a subset of the Teichmuller space of $\Sigma^2$ which is isomorphic to the Teichmuller space of $\Sigma$ (in particular, it is on the subset of $\mathscr{T}(\Sigma^2)$ which is invariant under $I$). Now we compile results, obtaining
\begin{equation}
\begin{aligned}
    W_{\Sigma} = \int_{\mathscr{F}} d\mu_{\mathrm{WP}, \Sigma} (R_{\Sigma^2, I}(J))^{\frac{-13}{2}}\ \left[
    (\det(P_1^{\dagger}P_1)_{g, \Sigma^2})^{\frac{1}{2}}\left(\frac{\det'(\Delta)_{g, \Sigma^2}}{ \int_{\Sigma^2}\sqrt{g}}\right)^{-13}
    \right]^{\frac{1}{2}}
    \  e^{-I(X_{cl},g)}. 
\end{aligned}
\end{equation}
At this stage we review our strategy to obtain an explicit representation of the partition function on $\Sigma$. We wish to write the above integrand in terms of moduli parameters on $\Sigma^2$ and restrict the integration to the subset of $\mathscr{T}(\Sigma^2)$ which is isomorphic to $\mathscr{T}(\Sigma)$ (in fact we restrict further to the moduli spaces). It is in fact easy to carry this out if one notes that the partition function on $\Sigma^2$ is given by \cite{Alvarez:1982zi}
\begin{equation}
    W_{\Sigma^2} =  \int_{\mathscr{F}_{\Sigma^2}}d\mu_{\mathrm{WP}, \Sigma^2}
    (\det(P_1^{\dagger}P_1)_{g, \Sigma^2})^{\frac{1}{2}}\left(\frac{2\pi}{ \int_{\Sigma^2}\sqrt{g}}\mathrm{det}'(\Delta)_{g, \Sigma^2}\right)^{-13} \ . 
\end{equation}
This is the 2-loop closed bosonic string partition function, and its value is well-known from the literature. We see that a detailed analysis of the above integrand in terms of a good set of moduli parameters, combined with knowledge of how the region of integration should be restricted, will allow us to immediately obtain $W_{\Sigma}$.

\subsection{Schottky Parametrisation}
The double torus can be represented as the complex plane (union infinity) with two pairs of non-intersecting circles cut out such that in each pair the two circles are isometric to each other; here, we give a summary of this representation as described, e.g.,  in \cite{TseytlinRG}. The handles on the double torus are formed by identification of the boundaries of each pair of circles. To make this identification, we associate with each pair of circles $(I_1,I'_1)$ and $(I_2,I'_2)$ a Mobius transformation $T_i:\mathbb{C} \rightarrow \mathbb{C} $ where
\begin{equation}
\begin{aligned}
&T_i(z) \equiv \frac{A_iz+B_i}{C_iz+D_i}, \hspace{0.5cm} A_i D_i - B_i C_i = 1, \hspace{0.5cm} i \in \{1,2\},\\
&T_i(I_i) = I'_i,
\end{aligned}
\end{equation}
i.e. $T_i$ maps the circle $I_i$ to the isometric circle $I'_i$. This leads to the fact that the isometric circles are given by 
\begin{equation}
    I_i = \{|C_iz+D|=1\}, \hspace{0.5cm} I'_i = \{|C_iz - A| = 1\}.
\end{equation}
The Schottky group $G$ is the group generated by these Mobius transformations. Then, the fundamental region of the double torus is the region of the complex plane exterior to the four circles $I_1$, $I'_1$, $I_2$ and $I'_2$. 

Furthermore, it can be shown that any double torus can be represented by the fundamental region of such a Schottky group. Since each of the Mobius transformations $T_i$ can be parametrised by three complex numbers $(\xi_i, \eta_i, k_i)$ where $k_i$ is the multiplier of $T_i$ and $\xi_i$ and $\eta_i$ are the repulsive and attractive fixed points of $T_i$, we have 6 complex parameters. However, if two sets of parameters are related by an overall $\mathrm{SL}(2,\mathbb{C})$ transformation, then the corresponding Riemann surfaces relate to the same point in moduli space. 

This reduces the number of complex parameters to 3 as expected. In fact, we will write the integrand in a $\mathrm{SL}(2,\mathbb{C})$ invariant form, and divide out by the volume of $\mathrm{SL}(2,\mathbb{C})$. Now we write down the partition function for the 2-loop closed bosonic string \cite{DiVecchia:1987uf} 
\begin{equation}\begin{aligned}
    W_{\Sigma^2} = &\frac{1}{\Omega}\int_{\mathrm{mod}(\Sigma^2)}\prod_{i=1}^{2}\left(\frac{d^2k_id^2\xi_i d^2\eta_i}{|k_i|^4|\eta_i-\xi_i|^4}|1-k_i|^4\right)(\det \  \mathrm{Im} \ \tau)^{-13}\\
    &\ \ \ \ \ \times  \prod'_{\alpha}\prod_{n=1}^{\infty}|1-k_{\alpha}^n|^{-48}\prod'_{\alpha}|1-k_{\alpha}|^{-4}, 
    \end{aligned}
\end{equation}
where $\prod_{\alpha}'$ is over all primitive elements of the Schottky group and $\Omega$ is the volume of the $\mathrm{SL}(2,\mathbb{C})$ group. We now assume that the Weil-Petersson measure in this expression is 
\begin{equation}
    d\mu_{\mathrm{WP}, \Sigma^2} = \prod_{i=1}^{2}\left(\frac{d^2k_id^2\xi_i d^2\eta_i}{|k_i|^4|\eta_i-\xi_i|^4}|1-k_i|^4\right).
\end{equation}

\subsubsection{Schottky Parametrisation of $\Sigma^2$}
We will express the disc-with-handle as the upper half complex plane with two isometric holes cut out. In this model, the boundary of the disc is given by the real axis, and the handle results from identification of the boundaries of the two circles. The opposite orientation copy is the lower complex plane with two isometric holes cut out. 

To be precise, the involution corresponds to complex conjugation. We form the Schottky double $\Sigma^2$ by attaching these two surfaces along the boundary, i.e. along the real axis, obtaining the complex plane with two pairs of isometric circles cut out, i.e. a double torus. In particular, we see that each point in the moduli space of $\Sigma$ corresponds to a point in moduli space of $\Sigma^2$ such that 
\begin{equation}
    I_1 = (I_2)^*, \hspace{0.5cm} T_1(z) = (T_2(z^*))^*, \hspace{0.5cm} k_1 \backslash \xi_1 \backslash \eta_1 = k_2^* \backslash \xi_2^* \backslash \eta_2^*.
    \label{1}
\end{equation}
\begin{figure}[H]
    \centering
    \includegraphics[scale=0.25]{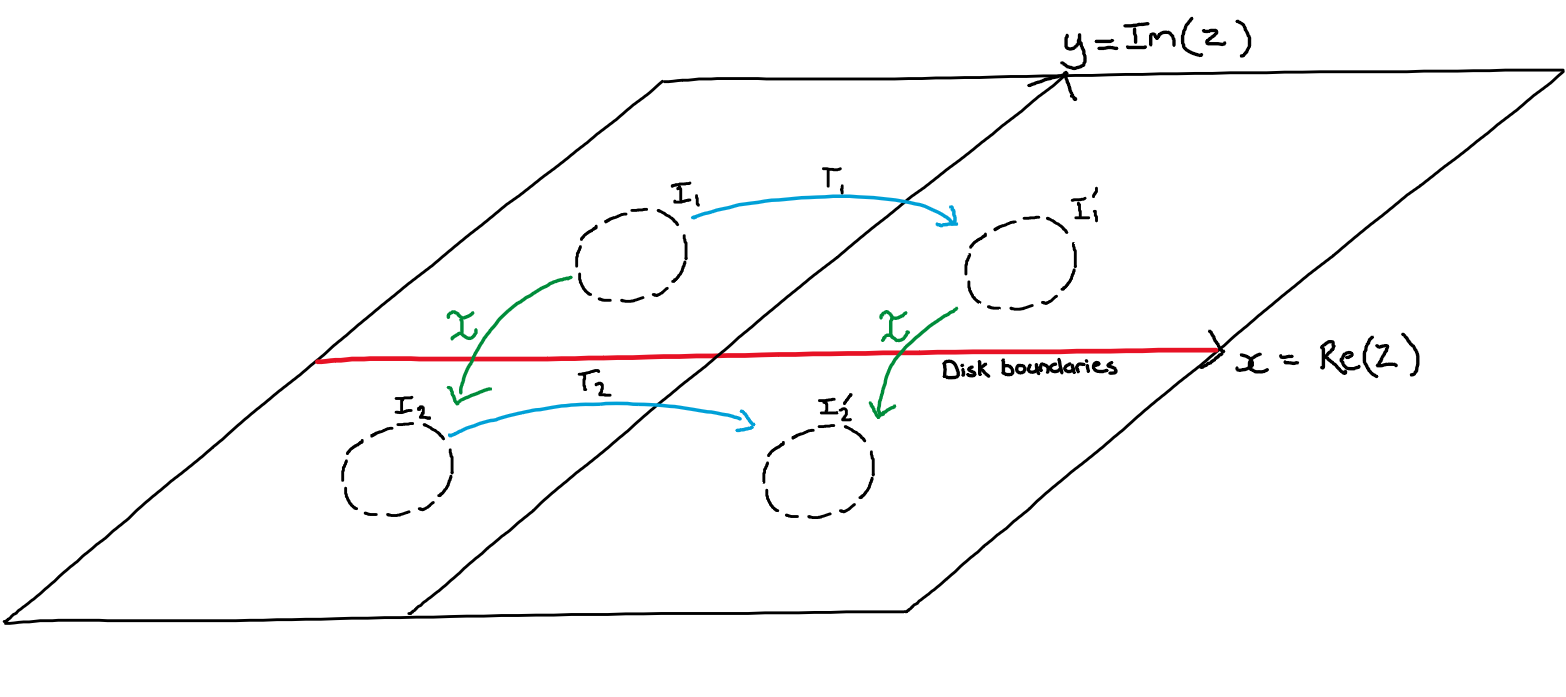}
    \caption{$\Sigma$ and its opposite orientation copy are related by the involution $I $ (shown in green), which is simply complex conjugation.}
    \label{}
\end{figure}
Thus, the Weil-Petersson measure for $\Sigma$ is given by 
\begin{equation}
   d\mu_{\mathrm{WP}, \Sigma} = \frac{d^2k_1d^2\xi_1 d^2\eta_1}{|k_1|^4|\eta_1-\xi_1|^4}|1-k_1|^4,
\end{equation}
whilst the integrand derived from the result for $\Sigma^2$ is simply restricted to the appropriate subset of moduli space. Thus, we obtain the result
\begin{equation}
\begin{aligned}
    W_{\Sigma} = &\frac{1}{\sqrt{\Omega}}\int_{\mathrm{mod}(\Sigma)}\frac{d^2k_1d^2\xi_1 d^2\eta_1}{|k_1|^4|\eta_1-\xi_1|^4}|1-k_1|^4(\frac{1}{2\pi}R_{\Sigma^2, I}(J))^{-\frac{13}{2}}\\
   &  \bigg((\det \  \mathrm{Im} \ \tau)^{-13} \ \prod'_{\alpha}\prod_{n=1}^{\infty}|1-k_{\alpha}^n|^{-48}\prod'_{\alpha}|1-k_{\alpha}|^{-4}\bigg)^{\frac{1}{2}}\ e^{-I_{\mathrm{cl}}(X,g)}\ ,
\end{aligned}
\end{equation}
where the term is brackets is calculated for the case where the relations in \ref{1} hold.

\subsection{Period Matrix Parametrisation}
Alternatively, the method in \cite{BLAU1988285} used to calculate the partition function of the three-holed sphere can be immediately adapted to the disc-with-handle. Here one uses the period matrix to parametrise the moduli space. Up to a constant independent of the metric one has \cite{BLAU1988285}
\begin{align}
    &d\mu_{\mathrm{WP}, \Sigma}     \left({(\det(P_1^{\dagger}P_1)_{g, \Sigma^2})^{\frac{1}{2}}}\left(\frac{\det'(\Delta)_{g, \Sigma^2}}{\det \ \textrm{Im} \ \tau \int_{\Sigma^2}\sqrt{g}}\right)^{-13}\right)^{\frac{1}{2}} \nonumber \\   
    = &\underbrace{\prod_{1\geq i\geq j\geq 2}\{\frac{1}{2}(1+I^*)d \ \mathrm{Im} \ \tau+\frac{1}{2}(1+I^*)d \  \mathrm{Re} \ \tau\}_{ij}}_{\text{Weil-Petersson Measure}} \prod_{\alpha \ \textrm{even}}|\theta^2\{\alpha\}(0,\tau)|^{-1}
\end{align}
where the second product is of the ten even genus two theta functions. Note that the Weil-Petersson measure for $\Sigma$ is given in terms of the imaginary and real parts of the period matrix $\tau$ of the doubled surface $\Sigma^2$. The integration, however, is over the subset of the Teichmuller space of $\Sigma^2$ which corresponds to the Teichmuller space of $\Sigma$. To calculate the Weil-Petersson measure, we use the fact that 
\begin{equation}
    I^* \tau = -\Gamma^{T}\Bar{\tau}\Gamma, \hspace{0.5 cm} \Gamma = \begin{pmatrix}
    0 & 1 \\
    1 & 0
    \end{pmatrix}.
\end{equation}
Note that the Teichmuller space of the double torus is the Siegel upper half-space with the subspace of diagonal period matrices removed \cite{Rodrigues:1986us}. Denoting $\tau$ as 
\begin{equation}
    \tau = \begin{pmatrix}
    a+ib & c+id \\
    c+id & e+if
    \end{pmatrix},
\end{equation}
we obtain
\begin{equation}
    \prod_{1\geq i\geq j\geq 2}\{\frac{1}{2}(1+I^*)d \ \mathrm{Im} \ \tau+\frac{1}{2}(1+I^*)d \  \mathrm{Re} \ \tau\}_{ij} = \frac{1}{2} dx \wedge dy \wedge dz, 
\end{equation}
where we have made a change of variables
\begin{equation}
    x = b+f, \ \ \ \  y = d, \ \ \ \ z = a-e.
\end{equation}
We find now that 
\begin{equation}
\begin{aligned}
    R_{\Sigma^2,I}(J) &= \frac{2}{bf -d^2}((b+d)^2+(d+f)^2)\\
    & = \frac{4x^2+16y(x+y)}{x^2 - (b-f)^2 -4y^2}.
\end{aligned}
\end{equation}
Compiling everything, 
\begin{equation}
    W_{\Sigma} \propto \frac{1}{\Omega'}\int dxdydz\left(x^2+4y(x+y)\right)^{-\frac{13}{2}}\prod_{\alpha \ \textrm{even}}|\theta^2\{\alpha\}(0,\tau)|^{-1} \ e^{-I_{\mathrm{cl}}(X,g)}.
\end{equation}
Note that we are integrating over all period matrices of the form 
\begin{equation}
    \tau = \begin{pmatrix}
    \frac{1}{2}(z+ix) & iy \\
    iy & \frac{1}{2}(z+ix)
    \end{pmatrix}
\end{equation}
subject to the condition that $\tau$ is in the Siegel upper half space with the subspace of diagonal period matrices removed \cite{Rodrigues:1986us}. We integrate over Teichmuller space and then `divide out' by the volume $\Omega'$ of the mapping class group of the disc-with-handle, which is the 3-stranded braid group. As before, the classical action is a function of the moduli, however, we do not evaluate it in this thesis. 

\section{Conclusion}
In this paper, we have reduced the infinite dimensional path integral  for the 
partition functions of the string world-sheets ending on a fixed circular  curve with annular topology and with the disc-with-handle topology to finite dimensional integrals over moduli parameters. For the annular topology, the partition function reduced to a one dimensional (divergent) integral over the inner radius of the annulus.  To do this  we calculated the determinant of the Laplacian operator on the annulus for the cases of Dirichlet-Dirichlet and Dirichlet-Neumann boundary conditions, which are factors that appear in the integrand.  

For the case of the disc-with-handle, we pursued two different methods in computing the partition function. In both cases, we utilised the Schottky double of the surface, that is the surface obtained by gluing the disc-with-handle to itself. This eliminates the problem of having to deal with a boundary on the surface. We then used the fact that determinants on a Schottky double are the square of Laplacian determinants on the single copy of the surface, and that we can write the Weil-Petersson measure on the single-copy surface via the Weil-Petersson measure on the doubled surface. We obtained the final result twice, first using Schottky parameters of the moduli space, and then using the period matrix.

\appendix

\acknowledgments
The authors would like to thank A. Tseytlin for taking them on for a summer project, which later developed into a Master's project. His continued guidance and support have been invaluable. We are also grateful for illuminating correspondence with R. Russo. The work of OS has been supported by the Theoretical Physics Undergraduate Summer Research Project funding. The work of BO has been supported by the Malcolm Weir foundation's summer funding.  




\bibliographystyle{JHEP}
\bibliography{FinalA.bib}

\end{document}